\documentclass[superscriptaddress,article,prl,showpacs,10pt,twocolumn,nofootinbib]{revtex4-1}
\usepackage{amsmath,xcolor,amsfonts,amsthm,amssymb,graphicx, srcltx,hyperref,ulem}
\usepackage{amsfonts}
\usepackage{amsmath}
\usepackage{epsfig}
\usepackage{xcolor,todonotes}
\usepackage{ulem} 
\usepackage{siunitx}
\newcommand{\beq}{\begin{equation}}
\newcommand{\eeq}{\end{equation}}

\newcommand{\musat}{\mu_{\rm sat}}

\renewcommand{\emph}[1]{{\it #1}}

\newcommand{\figref}[1]{Fig.~\ref{#1}}

\begin{document}
\title{Experimental single photon exchange along a space link of $7000~km$}
\author{Daniele Dequal}
\affiliation{Dipartimento di Ingegneria dell'Informazione, Universit\`a degli Studi di Padova, Padova, Italy}
\author{Giuseppe Vallone}
\affiliation{Dipartimento di Ingegneria dell'Informazione, Universit\`a degli Studi di Padova, Padova, Italy}
\affiliation{Istituto di Fotonica e Nanotecnologie, CNR, Padova, Italy}
\author{Davide Bacco}
\author{Simone Gaiarin}
\affiliation{Dipartimento di Ingegneria dell'Informazione, Universit\`a degli Studi di Padova, Padova, Italy}
\author{Vincenza Luceri}
\affiliation{e-GEOS spa, Matera, Italy}
\author{Giuseppe Bianco}
\affiliation{Matera Laser Ranging Observatory, Agenzia Spaziale Italiana, Matera, Italy}
\author{Paolo Villoresi}
\email{paolo.villoresi@dei.unipd.it}
\affiliation{Dipartimento di Ingegneria dell'Informazione, Universit\`a degli Studi di Padova, Padova, Italy}
\affiliation{Istituto di Fotonica e Nanotecnologie, CNR, Padova, Italy}
\date{\today}

\begin{abstract}
Extending the single photon transmission distance is a basic requirement for the implementation of quantum communication on a global scale. In this work we report the single photon exchange from a medium Earth orbit  satellite (MEO) at more than $7000~km$ of slant distance to the ground station at the Matera Laser Ranging Observatory. The single photon transmitter was realized by exploiting the corner cube retro-reflectors mounted on the LAGEOS-2 satellite. Long duration of data collection is possible with such altitude, up to 43 minutes in a single passage. The mean number of photons per pulse ($\musat$) has been limited to 1 for 200 seconds, resulting in an average detection rate of $3.0~counts/s$ and a signal to noise ratio of 1.5.
The feasibility of single photon exchange from MEO  satellites paves the way  to tests of Quantum Mechanics in moving frames and  to global Quantum Information. 
\end{abstract}
\maketitle

{\it Introduction -} Quantum Communications (QC) are necessary for tests on the foundation of Quantum Physics, such as Bell's inequalities violation \cite{Bell1964, Brunner2014, Hen2015}, entanglement swapping \cite{zuko93prl, xma2012} and distribution \cite{ina2013}, and quantum teleportation \cite{ma12nat}.
Moreover, the transmission of light quanta over long distances is crucial for the realization of several Quantum Information protocols, as quantum key distribution (QKD) \cite{benn84ieee, benn1999, hklo}, quantum authentication \cite{curty2001} and quantum digital signature \cite{wang2015}. 

One of the main challenges of QC consists in extending the length of the communication channel, which on ground is nowadays limited to few hundreds of kilometers, both for free space \cite{ma12nat} and optical fibers \cite{Korzh2015}. To overcome these limitations, in the last decade the feasibility of a quantum channel from an orbiting terminal has been explored both theoretically \cite{aspel2003, bona2006, toma2011, bour2013, schei2013, bruschiPRD2014} and experimentally 
by exploiting the Ajisai~\cite{vill08njp}  
and Champ~\cite{yin13ope} satellites.  Notably, recent transmission of qubits beyond Earth atmosphere \cite{vall2015} proved the feasibility of QC from 
several low Earth orbit (LEO) satellites, within $\num{1500}~km$ of altitude with respect to Earth surface and with typical passage duration of a few minutes. 
However, proposals for fundamental tests of Quantum Mechanics with moving frames, involving Special and General Relativity are based on link distances exceeding the LEO scale \cite{Ride2012, zych2012}. Moreover, satellites with orbits higher than LEO are  interesting  for the implementation of unconditionally-secure information protocols for  future global navigation satellite system (GNSS) constellations \cite{gerl2013} 
or for the realization of permanent links with
geostationary (GEO) satellites.

\begin{figure}[tt!!]
\centering
\includegraphics[width=8cm]{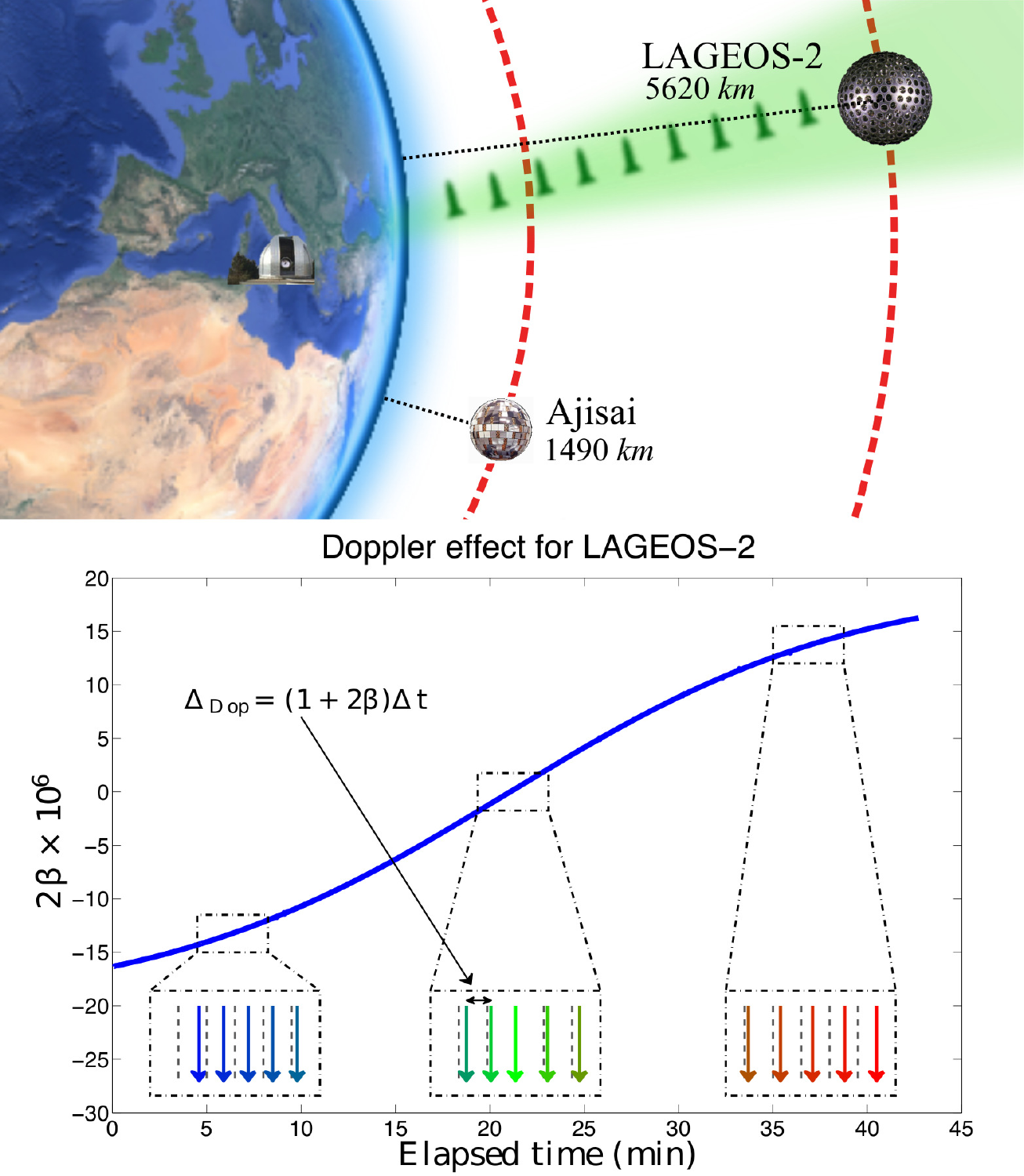}
\caption{(Color online) Top: schematic of the two-way protocol. A pulsed laser beam is directed toward a satellite equipped with CCRs and redirected toward the ground station. 
The two satellite trajectories are drawn to reproduce the distances in scale of Ajisai and LAGEOS-2 from the Earth surface. Bottom: measured  radial velocity $v_R$ in natural units $\beta=v_R/c$. 
The corresponding Doppler effect on the pulse separation is schematized on the three boxes: when $\beta<0$ ($\beta>0$) the pulse separations at the receiver are smaller (larger) than the transmitted pulse separations. The Doppler effect affects in the same way the $10~Hz$ SLR beam and the $100~MHz$ beam used for single photon generation.}
\label{fig:setup}
\end{figure}

In this perspective, this work aims at the expansion of QC with higher satellites by realizing the first single photon exchange spanning a distance exceeding $7000~km$, with a mean photon  number per pulse leaving the satellite ($\musat$) lower than 1 for an extended part of the link and with 
the single photon return signal above the noise.
Moreover we will
show that, with an upgrade of the detector on ground with commercially available parts, it would be possible to achieve signal to noise ratio (SNR) suitable for quantum communication up to $23000~km$, corresponding to the distance to GNSS.

{\it Experimental realization - } 
As sketched in the top panel of \figref{fig:setup}, 
to simulate a single photon source in orbit 
we exploited a train of light pulses directed towards the LAGEOS-2 satellite and
retroreflected by the satellite
corner cube retroreflectors (CCRs).
Such pulses are attenuated in the uplink by the combined effect of telescope divergence, optical diffraction,  atmospheric absorption and atmospheric turbulence \cite{degn93geo}. The mean photon number $\musat$ of the retroreflected pulses has been estimated by the link budget equation.
Because of CCRs backreflection capacity, 
the ground station is automatically in the illuminated cone on ground.

The pulses are directed towards
the satellite by a tracking telescope that is also used as single-photon receiver. 
The transmitter and receiver setups share the same telescope Coud\'e path  and are combined  by a beam splitter (BS).
The train of upgoing pulses is generated by  a mode-locking Nd:YVO$_4$ laser oscillator, operating at $1064~nm$ and with  repetition rate stabilized at $100~MHz$ by an atomic clock.
By means of a PPLN nonlinear crystal,   we obtained 
pulses at the wavelength of $532~nm$, with $1.1~nJ$ of energy and a duration of $100~ps$ FWHM. 
The beam size and divergence are corrected with a Galilean beam expander before entering the Coud\'e path of the telescope. 
Finally, the pulses are directed towards the satellite by the 
$1.5~m$  diameter primary mirror of  the Matera Laser Ranging Observatory (MLRO) of the Italian Space Agency, located in Matera, Italy.

The optical setup for the detection of the single photons from the satellite is composed by a focusing optics,  field-of-view (FOV) control, filtering  and a single photon detector, based on a photomultiplier tube (PMT), Hamamatsu H7360-02. The  PMT TTL output is time-tagged  with  a $81~ps$ time resolution by a fast time-to-digital converter (TDC), QuTAU from QuTOOLS. 
Two shutters are used to alternate the transmission and reception periods and to avoid the exposure of the PMT to stray upward light. 
The FWHM of the receiver FOV has been set to  $100~\mu rad$ for reducing the gathering of background. In addition, a  spectral filter of $3~nm$ FWHM transmission band around $532~nm$ has been used to reduced the stray light counts below the $50~counts/s$ intrinsic dark counts of the PMT.

{\it Temporal synchronization -} To discriminate  photons received from the satellite from the uncorrelated background, detected events are temporally filtered according to the instantaneous round trip time of light. Photon expected-times-of-arrival $t_{ref}$ at the receiver are not periodic during the satellite passage, as they depend on the actual distance from the ground station and on the radial velocity of the satellite. The time difference between two consecutive $t_{ref}$ deviates from the period of $10~ns$ imposed at the generation by a factor $ 1+2 v_R/c$, where $v_R$ is the satellite radial velocity and $c$ is the speed of light as shown in \figref{fig:setup}(bottom).

In \cite{vill08njp}, the $t_{ref}$'s
were obtained by calculating the satellite orbital parameters. Here, similarly to \cite{vall2015}, we have introduced an ancillary pulsed beam to simplify the 
prediction of $t_{ref}$.
Such ancillary beam realize the satellite laser ranging (SLR) in real time and it is derived from the same oscillator used for the generation of the single photon train, thus providing an intrinsic temporal 
synchronization between the two beams. 
The SLR beam is generated by a three stages laser amplifier system, and consists of a train of  $100~mJ$ laser pulses with a repetition rate of $10~Hz$. It is  combined together with the single photon beam toward and from the satellite. The exit and return instants of the SLR beam are detected using fast photodiodes and time-tagged by the same TDC used to register the PMT detections. Since both beams follow the same path to and from the satellite, the actual $t_{ref}$ of every pulse of the $100~MHz$ train  may be obtained in real time and with sub-ns accuracy 
by the Doppler effect measured on the SLR signal
(\figref{fig:setup} bottom).

\begin{figure}[t!]
\centering
\includegraphics[width=8cm]{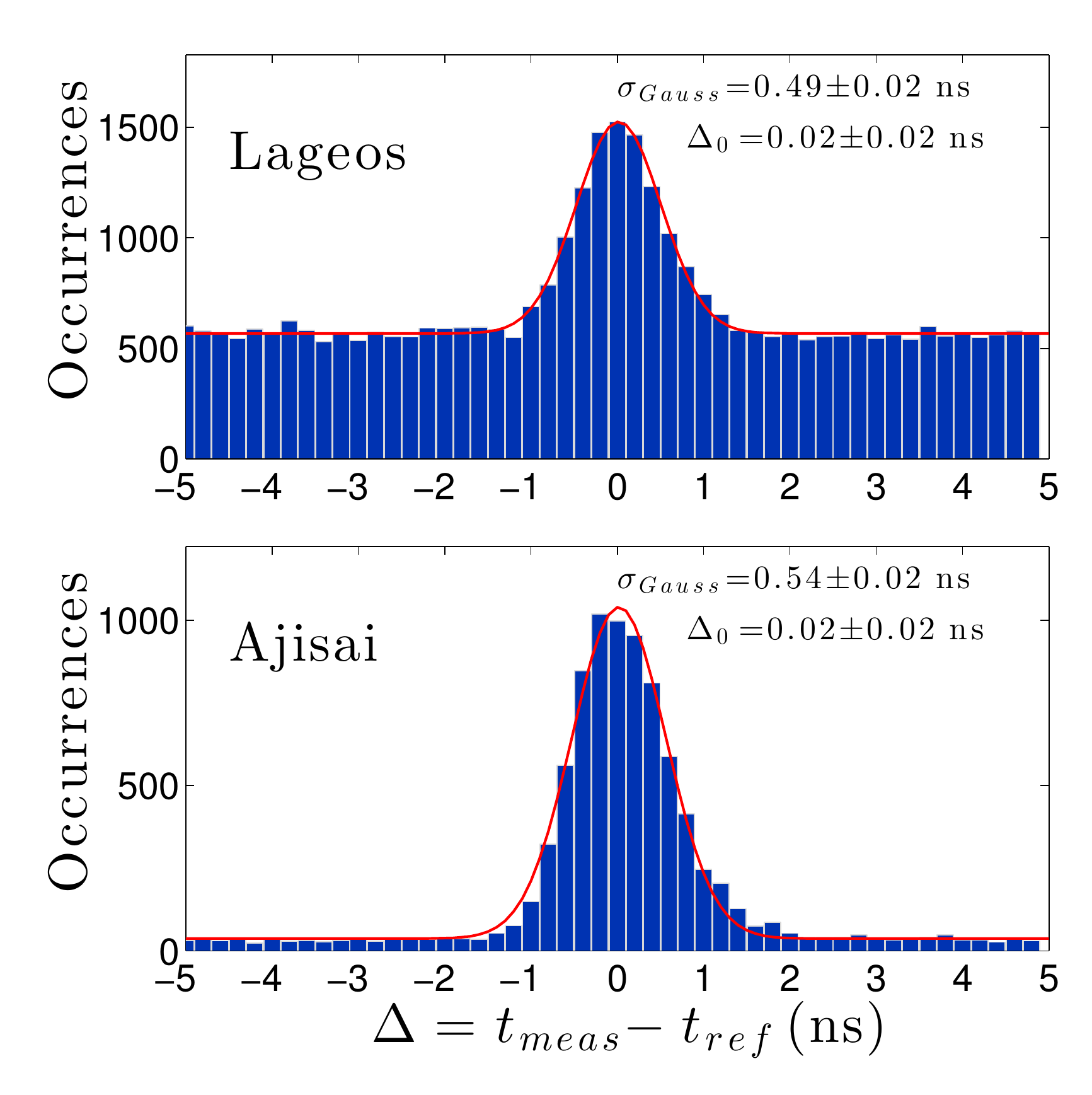}
\caption{(Color online) Comparison of the measured time of detection with the expected time of photon arrival. The compatibility of the fit parameters for LEO and MEO satellites demonstrates the robustness of the analysis method independently from the satellite distance.}
\label{fig:frequency}
\end{figure}

We tested the synchronization method  with two satellites in different orbits: Ajisai, a LEO satellite with altitude of $1490~km$ also used in \cite{vill08njp,vall2015}, and  LAGEOS-2, a MEO satellite with altitude of $5620~km$. For both satellites, we evaluated the difference of the measured time-tags of a complete passage with the nearest $t_{ref}$ calculated by the ancillary SLR pulses. The results are presented in Fig. 2 without any selection on the mean number of photons hitting the satellite: in both cases, a peak exceeding the background was obtained.

We extracted the residual offset ($\Delta_0$) and the standard deviation ($\sigma_G$) 
by  Gaussian fitting the distribution of the differences. 
The residual offset is $\Delta_0=0.02\pm0.02~ns$ in both cases
and results compatible with 0, thus demonstrating the validity of this synchronization technique for both the LEO and MEO satellites.
The measured standard deviations for the two passages are $0.49\pm0.02~ns$ for LAGEOS-2 and $0.54\pm0.02~ns$ for Ajisai. These two values are compatible within the experimental errors, indicating that the synchronization procedure can be effectively used also for larger satellite distance.
In both cases, the temporal precision of the single photon counts is essentially determined by the $0.50~ns$  jitter of the PMT detectors.

{\it Link budget analysis - }  The radar equation~\cite{degn93geo} predicts that the overall link losses, namely the ratio between the received ($f_{rx}$) and transmitted ($f_{tx}$) photon rate, depend on the satellite distance $R$ as follows:
\begin{equation}
\frac{f_{rx}}{f_{tx}} = \alpha \, G_t \, \frac{T_a^2(R)}{R^4}\,,
\label{eq:radar}
\end{equation}
where $\alpha=237~m^4$ is a coefficient that includes optical attenuation, geometrical factors, satellite cross section and detection efficiency for LAGEOS-2, $G_t $ is the transmission gain and $T_a$ the one-way atmospheric transmission coefficient.

In  \eqref{eq:radar}, the detection rate depends on the instantaneous satellite slant distance  $R$ both directly, due to the factor $1/R^4$, and implicitly in $T_a$, which is given by 
\begin{equation}
T_a(R) = T_0^{\frac{2 \, R \, (h_t+R_e)}
{(h_s+R_e)^2-R^2-(h_t+R_e)^2} \, 
\exp(-h_t/h_{\rm scale})}\,,
\label{eq:Tatm}
\end{equation}
where $T_0=0.89$ is the zenith transmission coefficient at sea level, 
$h_t = 537~m$ the observatory altitude above sea level, $R_e=6371~km$ the Earth radius, $h_s=5620~km$ the satellite altitude and $h_{\rm scale} = 1.2~km$ a scaling factor~\cite{degn93geo}.
We note that the slant distance $R$ is in general larger than the satellite altitude
show in \figref{fig:setup}.

We analyzed the entire passage of LAGEOS-2 by dividing it in intervals of $60~s$ each, in which the mean return rate  ($\bar{f}_{rx}$) was computed. To point out the events to be ascribed to the 100 MHz train from the uncorrelated background, we subtracted to the total detection rate the background rate measured from the events distant more than 6 $\sigma_{G}$ form $t_{ref}$, where $\sigma_{G}$ is the standard deviation of the Gaussian fit of \figref{fig:frequency}. 

\begin{figure}[tbp]
\centering
\includegraphics[width=8cm]{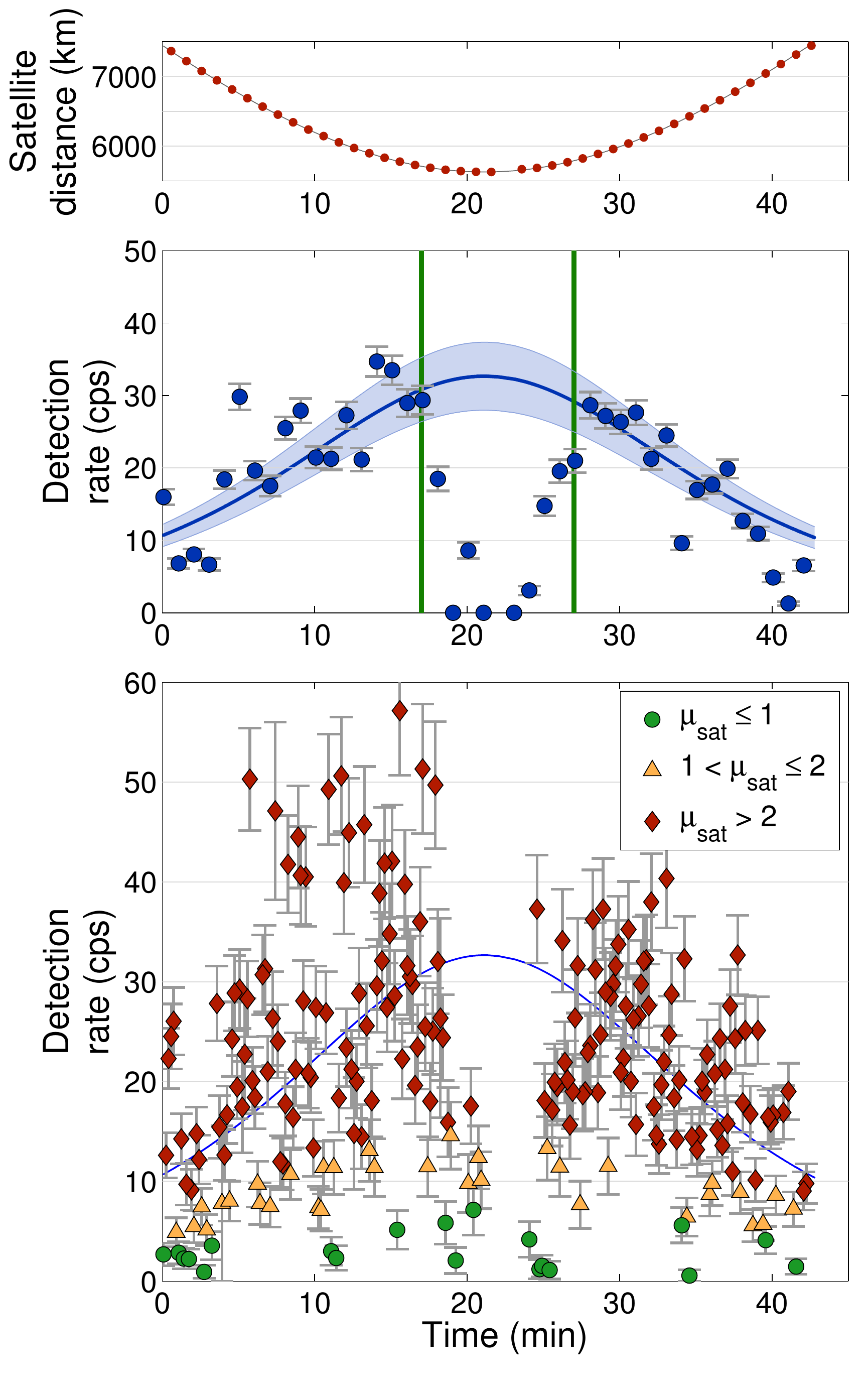}
\caption{(Color online) Top: satellite distance to the MLRO ground station, obtained with time of flight measurements of SLR pulses. Middle: single photon detection rate averaged over 60 sec compared with the global fit. The shaded area represent the 99.7\% of confidence level of the fit. Bottom: detection rate of single photons averaged over $10~sec$ 
for different values of $\musat$:
green circles, yellow circles
and red diamonds respectively 
correspond to $\musat \le 1$,  $1 < \musat \le 2$, and  $\musat > 2$.}
\label{fig:radar}
\end{figure}

\figref{fig:radar} shows the satellite distance (top) and the $\bar{f}_{rx}$ (middle) as a function of the elapsed time from the beginning of the satellite passage. It is possible to notice a drop in $\bar{f}_{rx}$ between minute 17 and 27. This is due to a larger telescope pointing error near the Zenith. With the exclusion of this interval, $\bar{f}_{rx}$ has been fit with \eqref{eq:radar}, leaving as free parameter the transmitter gain $G_t=\frac{8}{\theta_t^2} exp\left[-2\left(\frac{\theta_p}{\theta_t}\right)^2 \right]$, where $\theta_t$ is the fixed telescope divergence and $\theta_p$ is the time varying pointing error. The result of the fit for the whole passage of LAGEOS-2 is $G_t=(4.7 \pm 0.2) \times 10^8$. Excluding the above mentioned period, the total photon transmission time was then 32 min, that greatly extend the link durations previously reported \cite{vill08njp, yin13ope, vall2015}, that were at the level of few tens of seconds per satellite passage. This is due to two reasons: the duration of line-of-sight with MEO satellites is longer with respect to LEO, and the pointing accuracy is increased, due to the lower angular velocity of the satellite, and lower influence of gravitational perturbation of the Earth crust. 

{\it $\musat$ estimation -} The mean number of photons  hitting the satellite fluctuates due to turbulence and pointing errors. This effect is evident in the deviation of the measured detection rate from the global fit. To mitigate this effect, we based the assessment  of  $\musat$ not on the mean up-link attenuation but as the average number of received photon per pulse $\mu_{rec}$ divided by the down-link part of the radar equation. More specifically $\musat$ can be obtained from  
\begin{equation}
\mu_{rec} = \musat \, \frac{\Sigma}{A_{CCR} \, \rho \, N_{eff}} \dfrac{1}{4 \pi R^2} \,T_a\, A_t\, \eta_{rx}\, \,\eta_{det}\,,
\label{eq:downlink}
\end{equation}
where $A_{CCR}=11.4~cm^2$ \cite{arno1978} is the CCR reflective area, 
$\Sigma=15 km^2$ the Lageos cross-section,
$\rho=0.89$ \cite{mino1993} the CCR reflectivity at normal incidence, $A_t=1.7357~m^2$ the telescope area, $\eta_{rx}=0.1306$ the transmission of the whole receiving apparatus, 
$\eta_{det}=0.1$ the detector efficiency, and $N_{eff}=9.88$ \cite{arno1978} the effective number of CCR, averaged over all orientation. 

To reduce the effect of the fluctuation of the pointing error on $\musat$ estimation, we divided the whole passage in intervals of 10 s, calculating for each interval the mean $\mu_{rec}$. With this method we computed $\musat$ for each analysis slice, distinguishing those where $\musat \le 1$, $1 < \musat \le 2$, and $\musat > 2$ (\figref{fig:radar} bottom). The first two cases have a $\mu$ that can be considered practical to be used in protocols such as BB84 with decoy states \cite{tsuru2008, vall2015}. As shown in \figref{fig:radar}, we received pulses with less than one photon per pulse when the satellite is at the largest distance from the ground station,  corresponding to the maximum  attenuation, and when the telescope had pointing errors, which  increases the up-link losses.

\begin{figure}[t!]
\centering
\includegraphics[width=8cm]{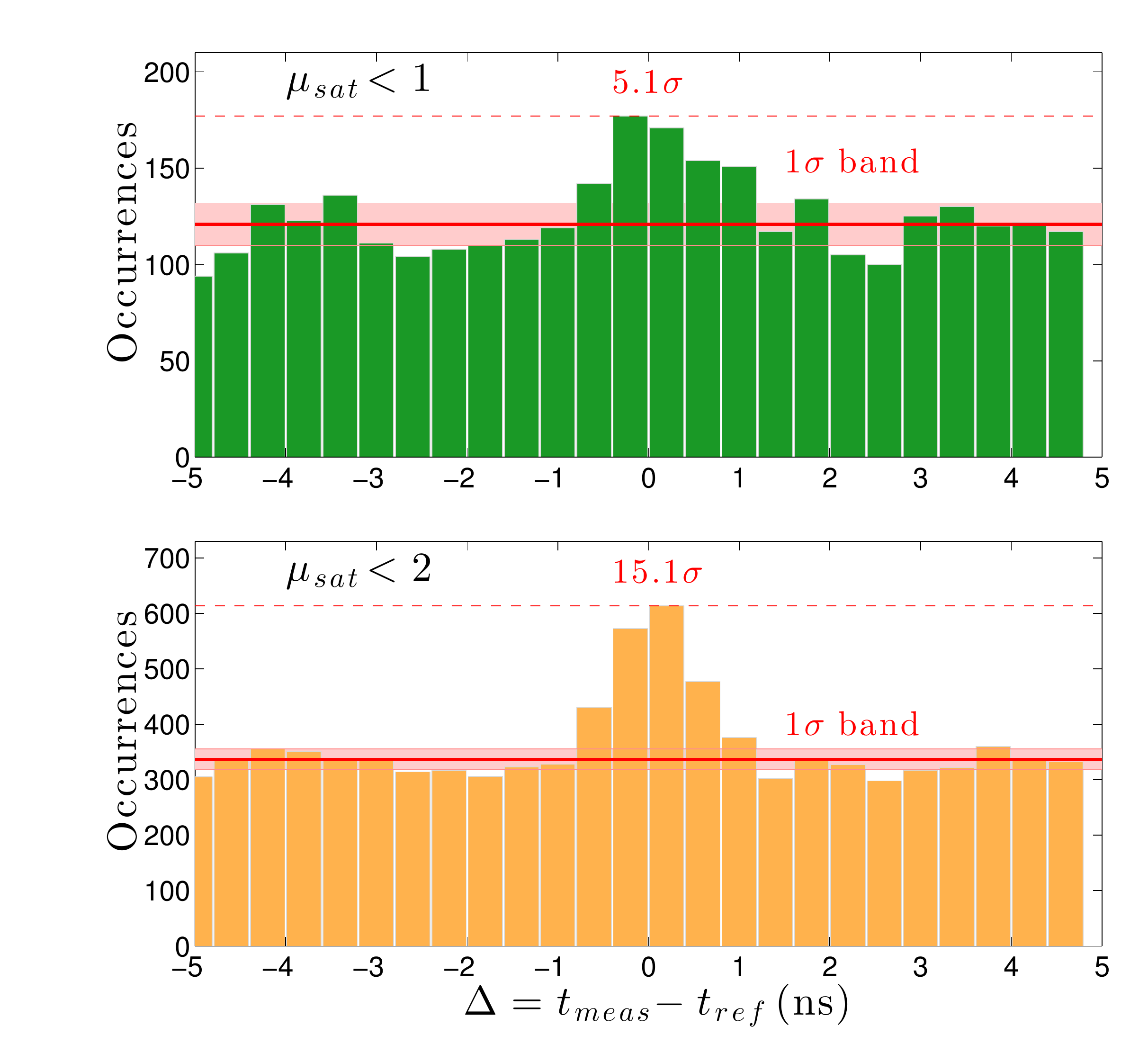}
\caption{(Color online) Histograms of detections as a function of deviation $\Delta$ from expected time $t_{\rm ref}$ for different values of $\musat$. 
When only intervals with $\musat<1$ are selected (top), a peak is observed at $\Delta=0$ with statistical significance over background of $5.1\sigma$. If the analysis is extended to $\musat<2 $ (bottom) the peak significance reaches $15.1\sigma$.
Bin size is $0.4~ns$. Robustness of the peaks has been verified by
changing the bin size.}
\label{fig:snr}
\end{figure}
By considering the intervals corresponding to $\musat \le 1$ it has been possible to obtain a total integration time of 200 s. For these periods we compared the timestamp of each photon detection with the nearest $t_{ref}$, with the same technique devised for the analysis reported in \figref{fig:frequency}. As shown in the upper graph of \figref{fig:snr}, we obtained a peak with statistical significance of  
$(5.1\pm1.2)\sigma$ over background (here $\sigma$ is the standard deviation of the background counts). The mean signal intensity for this intervals is $\musat=0.55\pm0.06 $ while the mean downlink attenuation is 72.3 dB. The average detection rate is $3.0\pm0.3~counts/s$, and the peak signal to noise ratio is SNR$=1.5\pm0.1$. 

We repeated the same analysis for intervals with $\musat \le $ 2 (\figref{fig:snr} bottom). This resulted in an integration period of $510~s$, a statistical significance of the peak over background of $(15.1\pm1.3)\sigma$, a mean $\musat=1.19\pm0.05 $ and a similar downlink attenuation, of 72.4 dB. The mean detection rate is $6.8\pm0.3$  counts/s and the peak ${\rm SNR}=1.8\pm0.1$. 

Based of these observations, we extrapolate the feasibility of QC with MEO satellites by simulating the photon link budget using the  best current technology of detectors  (silicon based single photon avalanche detectors) and under the hypothesis of effectively coupling on them the satellite returns. From Ref. \cite{giud2007},  peak detection efficiency is taken as $\eta_{S}=48\%$, FWHM timing accuracy as  $50~ps$ and dark count rate as ${\rm DC}_{S}\simeq350~counts/s$. These values significantly overpass the ones available for present experiment, $\eta_{PMT}=10\%$, ${\rm FWHM}_{PMT}=1.22~ns$ and ${\rm DC}_{PMT}=50~counts/s$, that are relative to the PMT technology, which was imposed by the current optical system coupled to the Coud\'e path.
We note that $50~ps$ are compatible with the expected temporal broadening of the pulse due to propagation, that are of the order of few $ps$ \cite{Marini73}.
Moreover, by adopting a modulated CCR as proposed in \cite{vall2015} with a light amplitude modulator allowing for $\mu_{fixed} = 0.6$ during the whole satellite passage, it would be possible to enhance the detection rate by a factor $\frac{\mu_{fixed}}{\mu_{sat}}\frac{\eta_{S}}{\eta_{PMT}}=5.2$ while reducing the dark count rate within the coincidence window by a factor $\frac{{\rm DC}_{PMT}}{{\rm DC}_{S}} \frac{{\rm FWHM}_{PMT}}{{\rm FWHM}_{S}}\simeq 3.45$. 

Because of these steps, the SNR increases by a factor 18 with respect to the values here obtained, corresponding to a quantum bit error ratio ${\rm QBER}\simeq 3.6 \% $. In the hypothesis that no other effects would step in the budget \cite{baea2007}, QKD from a MEO satellite appears as feasible. 
For extending QC to higher orbit such as the GNSS,  between $19000~km$ and $23000~km$ of altitude, a possible improvement of the detector dark counts down to $\simeq100$ counts/s is necessary. In this case the signal would be reduced by a factor 10.8, due to the $1/R^2$ factor in \eqref{eq:downlink}, giving a value of ${\rm QBER}\simeq6.6 \%$ that is compatible with the implementation of QKD for such GNSS satellites.

A possible improvement could come from the use of superconducting single photon detectors. With nanowire superconducting technology it is possible to obtain efficiency up to $80\%$, dark count rates as low as 10 Hz and time jitter of 40 ps. With these parameters it would be possible to obtain  a ${\rm QBER}=0.5\%$ form a satellite with the same orbital parameters as LAGEOS, and ${\rm QBER}=3\%$ from a GNSS satellite. However it mus be noticed that the smaller sensitive area of superconductive devices, which reaches $\sim 16\mu m$ compared to $\sim 200 \mu m$ of silicon based detectors, might lead to a lower efficiency due to the optical coupling.

{\it Conclusions -} We report on the experimental exchange of single photons from the LAGEOS-2 satellite to the MLRO ground station, with a link length exceeding $7000~km$.  The photon transmission has been kept stable for most of the passage of the MEO satellite, allowing for the assessment of the signal attenuation due to beam divergence and telescope pointing error by means of the established link models. 
By using an active source on satellite and commercially
available silicon detectors, a QBER suitable for the implementation of a QKD link from a MEO satellite appears as feasible. This also paves the way to endeavoring QC with GEO satellites and eventually to fundamental tests of combined Quantum Mechanics and General Relativity.

\acknowledgements

We would like to thank Francesco Schiavone, Giuseppe Nicoletti, and the MRLO technical operators for the collaboration and support. 
Our work was supported by the Strategic-Research-Project QUINTET of the Department of Information Engineering, University of Padova, the Strategic-Research-Project QUANTUMFUTURE of the University of Padova.

\end{document}